\documentclass[12pt]{article}
\usepackage{epsfig}
\usepackage{psfrag}
\usepackage{enumitem}
\usepackage{latexsym}
\usepackage{indentfirst}
\usepackage{fancyhdr}
\usepackage{amssymb}
\usepackage{amsmath}
\usepackage{amsfonts}
\usepackage{pifont}
\usepackage{cite}
\usepackage{bbold}
\usepackage{color}
\usepackage{slashed}
\usepackage{graphics}
\usepackage[center,footnotesize,hang]{subfigure}
\usepackage{url}
\usepackage{array}

\def\be{\begin{equation}}
\def\ee{\end{equation}}
\def\bc{\begin{center}}
\def\ec{\end{center}}
\def\bea{\begin{eqnarray}}
\def\eea{\end{eqnarray}}
\newcommand{\bi}{\begin{itemize}}
\newcommand{\ei}{\end{itemize}}                 

\newcommand{\ba}{\begin{array}{c}}
\newcommand{\bad}{\begin{array}{ccc}}
\newcommand{\ea}{\end{array}}

\def\nn{\nonumber}

\begin{document}

\begin{titlepage}
\hfill{RM3-TH/12-4}
  \vskip 2.5cm
  \begin{center}
   {\Large\bf  $S_3$ as a flavour symmetry for quarks and leptons after the Daya Bay result on $\theta_{13}$}
   \end{center}
  \vskip 0.2  cm
  \vskip 0.5  cm
   \begin{center}
   {\large Davide Meloni }~\footnote{e-mail address: davide.meloni@fis.uniroma3.it}
   \\
   \vskip .2cm {\it Dipartimento di Fisica "E. Amaldi"}
   \\
   {\it Universit\'a degli Studi Roma Tre, Via della Vasca Navale 84, 00146 Roma, Italy}
   \\
   \end{center}
  \vskip 0.7cm
 \begin{abstract}
 \noindent
We present a model based on the flavour group
$S_3 \times Z_3 \times Z_6$ to explain the main features of 
fermion masses and mixing. In particular, in 
the neutrino sector the breaking of the $S_3$ symmetry 
is responsible for a naturally small $r=\Delta m^2_{sol}/\Delta m^2_{atm}$ and suitable next-to-leading order corrections bring $\theta_{13}$ at the level 
of $\sim 0.13$, fully compatible with the recent Daya Bay result.
%
In the quark sector, the model accommodates the different mass hierarchies in the up and down quark sectors
as well as the Cabibbo angle and  $|V_{cb}|$ (or $|V_{ub}|$, 
depending on the charge assignment of the right-handed b-quark) in the correct range.
\end{abstract}
\end{titlepage}

\section{Introduction}
One of major challenges in the model building is to provide a framework where 
the hierarchies in the fermion masses and the values of the mixing angles can be explained 
in a natural way. Recent results in the neutrino sector \cite{Fogli:2011qn} have inspired many
authors to discuss the possible origin of a relatively large reactor angle, as 
obtained first by the T2K and MINOS collaborations \cite{Abe:2011sj} and, more recently, by 
Double Chooz \cite{Abe:2011fz} and Daya Bay \cite{db}. In particular, 
the latter has given a 5.2$\sigma$ evidence of $\theta_{13}\ne 0$:
\begin{eqnarray}
\label{res}
 \sin^2 2\theta_{13} = 0.092 \pm 0.016 \pm 0.005\,,
\end{eqnarray}
probably solving the longstanding question of the magnitude of $\theta_{13}$ (see \cite{Xing:2012ej},\cite{meloni},\cite{Branco:2012vs},
\cite{Zhang:2012ys} and \cite{He:2012yt} for a discussion of the impact of
such a result for lepton flavour mixing and leptonic CP violation).
Several models based on the breaking of additional symmetries beyond the Standard Model have been
proposed to explain this and other peculiarities emerged from neutrino oscillation experiments, as 
the large solar and atmospheric mixing angles as well as a small solar-to-atmospheric mass differences ratio $r$.
In an unified view of fermion mass hierarchies and flavour mixing, the quark sector must also be 
taken into account and this poses the question of how to reconcile the numerous differences 
in the spectra and mixing of quarks compared to leptons, the most relevant ones being 
a different structure of the CKM mixing matrix and the strong mass hierarchy in the up-quark 
sector.
 Although many attempts have been done in this 
direction, the question of explaining the values of fermion masses and mixing (one of the aspects of the flavour problem) 
is still an open issue and deserves further studies.
Discrete symmetries \cite{Altarelli:2010gt} have been invoked as a powerful tool 
to solve/mitigate some of these problems and, among several choices, the non-abelian discrete group $S_3$ \cite{Haba:2005ds} 
has also been investigated. Remarkably, models based on $S_3$ describing both quarks and leptons
(with very few assumptions on the scalar and Yukawa sectors of the theory) are not very common. 
Interesting attempts in this direction have been done in the context of GUT theories \cite{Koide:2003rx} 
and in a non-unified approach \cite{Chen:2004rr}-\cite{Feruglio:2007hi}.
In this paper we want to contribute to this discussion, presenting a SUSY see-saw model based on the 
flavour group $S_3 \times Z_3 \times Z_6$. Identifying $\varepsilon$ as the small $S_3$ breaking parameter (whose magnitude will be discussed 
later), the main features of our construction are:
\bi
\item the neutrino mass hierarchy can only be of normal type;
\item the solar-to-atmospheric mass differences ratio $r$ is naturally small, at the level 
of ${\cal O}(\varepsilon^2)$, without invoking any fine-tuning
among the Yukawa parameters (Sect.\ref{sectLO});
\item after the inclusion of suitable next-to-leading order corrections (NLO), the 
vanishing leading order (LO) $\theta_{13}$ receives a relatively large shift of ${\cal O}(\varepsilon)$ whereas 
$\theta_{23}$ is only modified by terms of ${\cal O}(\varepsilon^2)$; the solar angle remains large,
of ${\cal O}(1)$ (Sects.\ref{nlosect}-\ref{numan});
\item in the quark sector, we are able to reproduce the correct order of magnitude of the 
mass hierarchies in both up and down sectors (Sect.\ref{quarksect});
\item the Cabibbo angle is predicted in the correct range as well as $|V_{cb}|$ (or $|V_{ub}|$, 
depending on the charge assignment of the right-handed b-quark);
\item the flavon alignments needed to get the previous results 
are obtained from the minimization of the superpotential allowed by the symmetry of the model (Sect.\ref{flavonalignment}).
\ei
A bit more problematic is the explanation of the muon-to-tau mass ratio and the value of  $|V_{ub}|$
(or $|V_{cb}|$), in any case naturally smaller than the Cabibbo angle. 
Other interesting features of our model are the prediction for the effective mass $|m_{ee}|$,
quite small as usual for neutrinos with normal mass ordering, and the fact that
the Tri-bimaximal mixing structure (TBM) \cite{Harrison:2002er} can be easily obtained using
only one additional relation among the Yukawas (not dictated by the symmetry of the theory).

The model is formulated in the framework of the SUSY see-saw mechanism. 
At LO, the contribution to $\theta_{12}$ only comes from
the neutrino sector whereas  $\theta_{23}$ is completely generated by the charged lepton sector. 
At this order, the reactor angle is vanishing. 
To produce such a pattern, the mass matrices $m_\ell\,m^\dagger_\ell$ and $m_\nu$ must be block-diagonal in the (23)
and (12) sectors, respectively. In particular, having a block-diagonal light neutrino mass matrix
helps in giving different order of magnitude to the mass eigenstates and then in obtaining a small $r$.

We use a {\it non-conventional} assignment of the left-handed doublets $L$ of the second and third generations in 
a 2 representation of $S_3$ as follows:
\be
L=\left(
\begin{array}{c}
\tau \\
\mu
\end{array}
\right)\,,
\ee
whereas for the right-handed doublet we assume:
\be
L^c=\left(
\begin{array}{c}
\mu^c \\
\tau^c
\end{array}
\right)\,.
\ee
Electron fields are assigned to the singlet $1$.
In the neutrino sector, we introduce three right-handed fields:
the first two generations are grouped in the 2 representation:
\be
\nu^c=\left(
\begin{array}{c}
\nu_e^c \\
\nu_\mu^c
\end{array}
\right)\,.
\ee
whereas $\nu_3^c$ belongs to a $1_2$.
We use the following representation of the $S_3$ group:
\be
S^2=T^3=(ST)^2=1~.
\ee
In the so-called ``T-diagonal`` basis we have:  
\begin{equation} 
\label{bidimofst}
S = \left(
\begin{array}{cc} 
0 & 1 \\
1 & 0 
\end{array}
\right)
\qquad     
T = \left( 
\begin{array}{cc} 
\omega & 0 \\
0 & \omega^2 \\ 
\end{array}   
\right) , 
\end{equation}
with $\omega=e^{i 2/3 \pi}$.
The tensor products involving pseudo-singlets are given 
by $1_2\times 1_2 =1 $ and $1_2 \times 2=2$ while the product of two doublets is  
$2\times 2=2+1+1_2$ which, in terms of the components  of the two doublets $A=(a_1,a_2)^{\rm T}$ 
and $B=(b_1,b_2)^{\rm T}$, are as follows:
\be 
\label{tensorprod1s3}
a_1b_2+a_2b_1 \in 1 
\qquad
a_1b_2-a_2b_1 \in 1_2 
\qquad
\left( 
\ba{}
a_2 b_2 \\
a_1 b_1 \\
  \ea
  \right)
  \in 2\,.
 \ee
To ensure the breaking of $S_3$ along appropriate directions in the flavour space, we need 
two doublet flavons $\phi$ and $\xi$ with the following vevs (derived from the minimization of the superpotential, see Sect.\ref{flavonalignment}):
\be
\label{loflav}
\langle \phi\rangle= v_\phi\,
\left(
\begin{array}{c}
1 \\
1
\end{array}
\right) \qquad 
\langle \xi\rangle= v_\xi\,
\left(
\begin{array}{c}
0 \\
1
\end{array}
\right) \,;
\ee
two other scalar singlets $\chi\sim 1$ and $\chi^\prime\sim 1_2$ are also needed to guarantee  appropriate non-vanishing 
entries in the fermion mass matrices;  both singlets have non vanishing vevs:
\bea
\label{loflav2}
\langle \chi\rangle= v_\chi \qquad \langle \chi^\prime\rangle= v_\chi^\prime\,.
\eea
The particle content and the transformation properties of leptons, electroweak Higgs doublets and flavons under $S_3 \times Z_3 \times Z_6$
are summarized in Tab.\ref{transform}.
\begin{table} [h!]
\centering
\begin{tabular}{|c||c|c|c|c|c|c||c|c|c|c|c||c|c|c|}
\hline
{\tt Field}& $\nu^c$ & $\nu^c_3$ & $L_e$ & $L_e^c$ & $L$ & $L^c$ & $h_{u,d}$& 
$\phi$ &  $\chi$ & $\xi$ & $\chi^\prime$ & $\xi_0$  & $\Delta_0$  \\
\hline
$S_3$ & $2$ & $1_2$ & $1$ & $1$ & $2$ & $2$ &$1$ &$2$ & $1$ & $2$ & $1_2$ &  1 &  $2$  \\
\hline
$Z_6$ & $\omega$ & $\omega$ & $1$ & $\omega^3$ & $\omega^5$ & $\omega^3$ & 1 &
$\omega^4$ & $\omega^4$  & $\omega^4$ & $\omega^5$  & $\omega^4$ &  $\omega^4$  \\
\hline
$Z_3$ & 1 & 1 & $1$ & $\omega$ & 1 & $\omega^2$ & 1 &
$\omega$ &$\omega$  & $1$ & $1$ & 1 &  $\omega$\\
\hline
$U(1)_R$ & $1$& $1$ & $1$ & $1$ & $1$ & $1$ & $0$ & $0$& $0$  & $0$ & $0$ & $2$ & $2$  \\
\hline
\end{tabular}
\caption{\it Transformation properties of leptons, electroweak Higgs doublets and flavons under $S_3 \times Z_3 \times Z_6$ and $U(1)_R$~.}
\label{transform}
\end{table}

\section{Leptons: leading order}
\label{sectLO}
\subsection{Charged leptons}
The leading order lagrangian in the left-right basis reads:
\bea
{\cal L}_{cl} = x_1\, L\, L^c\, h_d\, \frac{\phi}{\Lambda} +x_2\, L\, L^c\, h_d\, \frac{\chi}{\Lambda} \,.
\eea
At this stage, there are no contributions to the electron entries (since $\chi^\prime$ has been assigned to the $1_2$ representation)
and the electron mass is vanishing.
In the $\mu-\tau$ subsector the mass matrix is as follows:
\be 
\label{chargedLO}
m_\ell= \frac{v_d}{\Lambda}\,
\left(\begin{array}{cc}
x_2 v_\chi& x_1 v_\phi\cr
x_1 v_\phi&x_2 v_\chi
\end{array}\right)\,,
\ee
where $v_d = \langle h_d \rangle$.
The phases of the Yukawas $x_{1,2}$ can be rotated away, with no loss of generality. 
The charged lepton masses and the mixing matrix are then:
\be
m_\mu=  \frac{v_d}{\Lambda} \left|x_1 v_\phi +x_2 v_\chi\right| \qquad 
m_\tau =  \frac{v_d}{\Lambda} \left|x_1 v_\phi -x_2 v_\chi\right|\,,
\ee
\be 
U_\ell=\frac{1}{\sqrt{2}}\,
\left(\begin{array}{cc}
1 & -1 \cr
1  & 1
\end{array}\right)\,.
 \ee
The question is how to realize a naturally small muon-to-tau mass ratio. In \cite{grimus} it has been suggested to introduce a further 
symmetry under which $x_1 = -x_2$ and $v_\phi = v_\chi$; however, as pointed out in \cite{Feruglio:2007hi}, this additional
symmetry does not commute with $S_3$ and the whole symmetry group is larger than the latter.
Here we observe that for real flavon vevs (as assumed here)
any couple of real $(x_1,x_2)$ determines a larger and a smaller eigenvalue, so a mass splitting is quite natural
in this model. If we want to correctly reproduce the mu-to-tau mass ratio we have to fine-tune the Yukawa parameters;
in particular, assuming $m_\mu/m_\tau \sim \lambda^2$ ($\lambda$ being the Cabibbo angle) we get:
\bea
x_1 = - x_2 \,(1 + 2 \,\lambda^2) + {\cal O}(\lambda^4)\,,
\eea
(so that, as expected, the two Yukawas must be opposite in sign and almost equal in magnitude). 
We notice that, at this level, we do not need to specify a value of the ratios of the flavon vevs over the cut-off $\Lambda$.

\subsection{Neutrinos}
\label{subnu}
The first useful terms to generate a Dirac mass matrix are:
\be
\label{diraclag}
{\cal L}_D = a\, L\, \nu^c\,h_u + b\, L_e \,\nu_3^c \,h_u\,\frac{\chi^\prime}{\Lambda}\,,
\ee
so that:
\be 
\label{diracmass}
m_D= v_u\,
\left(\begin{array}{ccc}
 0& 0 & b\,\frac{\chi^\prime}{\Lambda} \cr
a &0 & 0\cr
0&a&0
\end{array}\right)\,,
\ee
where $v_u = \langle h_u \rangle$.
For the Majorana mass matrix we have:
\be
\label{majolag}
{\cal L}_M = \alpha \,  \nu^c\,\nu^c\,\xi + \beta\,  \nu^c\,\nu_3^c\,\xi + \gamma\,\nu^c\,\nu^c\,\frac{{\chi^\prime}^2}{\Lambda} + 
\delta\,\nu_3^c\,\nu_3^c\,\frac{{\chi^\prime}^2}{\Lambda}
\,,
\ee
and:
\be 
\label{majo}
m_M= 
\left(\begin{array}{ccc}
 0& \gamma\,\frac{  v_\chi^{\prime\,2}}{\Lambda} & \beta\,v_\xi \cr
 \gamma\,\frac{v_\chi^{\prime\,2}}{\Lambda} &\alpha \,v_\xi& 0\cr
\beta\,v_\xi& 0& \delta\,\frac{v_\chi^{\prime\,2}}{\Lambda}
\end{array}\right)\,.
\ee
The first contribution to the heavy masses arise from the previous matrix with $v_\chi^\prime = 0$ and are:
\bea
\alpha\,v_\xi\,, \;\pm  \beta\,v_\xi\,.
\eea
The degeneracy is then lifted by taking into account the corrections from the $v_\chi^\prime$ terms.
The light neutrino mass matrix is obtained from the see-saw formula:
\be
\label{seesaw}
m_\nu = - m_D\,m_M^{-1}\,m_D^T\,;
\ee
to make the following analytical evaluations more readable, we fix $a=1$ (but it will be considered as a free parameter in the numerical evaluations to follow) 
and assume an equal order of magnitude of the 
flavon vevs to the cut-off ratio $\frac{\langle \Phi \rangle}{\Lambda}=\frac{v_\Phi}{\Lambda}=\varepsilon$, that we take as a small parameter.
At first order in $\varepsilon$ we get:
\be 
\label{massmu}
m_\nu= -\frac{v_u^2}{v_\Phi}\,
\left(\begin{array}{ccc}
 0& \frac{b}{\beta}\,\varepsilon  & 0 \cr
\frac{b}{\beta}\,\varepsilon & -\frac{\delta}{\beta^2}\,\varepsilon & 0\cr
0& 0&\frac{1}{\alpha}
\end{array}\right)\,.
\ee
All phases can be reabsorbed into a redefinition of the right-handed neutrino fields, so we deal with real parameters. We 
see that, at this order, we get a block-diagonal form of the light mass matrix, with the contribution to the (12) sector of ${\cal O}(\varepsilon)$
and the (33) element larger.
The matrix $m_\nu$ is compatible with normal hierarchy only and the light masses (still at the first order in $\varepsilon$) are then given by:
\bea
|m_1| &=&  \left(\frac{v_u^2}{v_\Phi}\right)\,\left[\frac{ (4 b^2 \beta ^2+\delta ^2)^{1/2}-\delta}{2 \beta ^2}\right]\,\varepsilon  \nn\\
&& \nn \\
|m_2| &=&  \left(\frac{v_u^2}{v_\Phi}\right)\,\left[\frac{(4 b^2 \beta ^2+\delta ^2)^{1/2}+\delta}{2 \beta ^2}\right]\,\varepsilon \label{massesLO}
\\
&& \nn \\
|m_3| &=& \left(\frac{v_u^2}{v_\Phi}\right)\,\frac{1}{\alpha}\nn
\eea
so that the solar and atmospheric mass differences read:
\bea
\Delta m^2_{sol} &=&  \left(\frac{v_u^2}{v_\Phi}\right)^2\,\frac{\delta (4 b^2 \beta ^2+\delta ^2)^{1/2}}{\beta ^4}\,\varepsilon ^2 \nn \\
&&\label{solatmLO}  \\
\Delta m^2_{atm} &=& \left(\frac{v_u^2}{v_\Phi}\right)^2\,\,\frac{1}{\alpha^2} + {\cal O}(\varepsilon^2) \nn\,.
\eea
Taking $\alpha = 1$, $\Delta m^2_{atm}=2.5 \times 10^{-3}$ eV$^2$ and $v_u = 100$ GeV,  we estimate $v_\Phi \sim 2\times 10^{14}$ GeV, which is a 
common order of magnitude for the heavy neutrino masses.  
It is now easy to  derive the ratio $r$:
\bea
r &=& \frac{\Delta m^2_{sol}}{\Delta m^2_{atm}} =\frac{\alpha^2\,\delta\, (4 b^2 \beta ^2+\delta ^2)^{1/2}}{\beta ^4}\,\varepsilon ^2  
\eea
which, for ${\cal O}(1)$ parameters, is naturally suppressed by $\varepsilon^2$. 
The value of $\varepsilon$ cannot be precisely determined at this stage;  
to recover the experimental value $r\sim 1/30$, 
$\varepsilon$ should be as small as the Cabibbo angle but a small numerical enhancement (suppression) 
of the coefficient $\frac{\alpha^2\,\delta\,(4 b^2 \beta ^2+\delta ^2)^{1/2}}{\beta ^4}$
can bring it to smaller (larger) values.
The matrix in eq.(\ref{massmu}) determines a non-vanishing solar angle: a straightforward computation gives:
\bea
\tan^2 \theta_{12} = 1-\frac{2 \delta }{(4 b^2 \beta ^2+\delta ^2)^{1/2}+\delta }\,,
\eea
so that it is generically of ${\cal O}(1)$. 
Summarizing, the whole neutrino mixing matrix at LO $U_{PMNS} = U_\ell^\dagger\,U_\nu$ has a vanishing reactor angle, maximal $\theta_{23}$ and large  ${\cal O}(1)$ solar angle.

It is interesting to observe that, to reproduce the TBM value $\tan^2 \theta_{12} =1/2$, one
simply needs
\bea
\label{tbm}
\delta=\pm b\,\beta/\sqrt{2}\,, 
\eea
which is still a number of ${\cal O}(1)$;
in this respect, this external condition does not appear to be completely unnatural since 
it does not require any strong hierarchy among the model parameters. Another interesting 
possibility to get the TBM matrix from the $S_3$ symmetry has been proposed in \cite{Mohapatra:2006pu} where, however,
after the inclusion of the charged lepton corrections one 
of the two allowed $S_3$ invariant Yukawa couplings must be switched off by hand.
Notice that a maximal value for $\theta_{12}$ in our model can only be obtained if $\delta=0$, which implies a vanishing 
$\Delta m^2_{sol}$. 

\section{Flavon alignment}
\label{flavonalignment}
The structure of the flavon vevs can be obtained minimizing the scalar superpotential in the limit of exact SUSY \cite{Altarelli:2005yx}.
Within this approach, a  continuous $U(1)_R$ symmetry is introduced, under which matter fields have $R=1$, while
Higgses and flavon fields have $R=0$. Such a symmetry will be
eventually broken down to the R-parity by small SUSY breaking
effects which can be neglected in the first approximation. Since the superpotential must have $R=2$, we need to
introduce two additional scalar fields, a doublet $\Delta_0$ and a singlet $\xi_0$, with 
$R=2$. Within these assumptions, the relevant part of the scalar potential of the model is given by
 the F-terms
with: 
\be
V_F=\sum_i\left\vert\frac{\partial w}{\partial\varphi_i}\right\vert^2\,.
\ee
In the following, we parametrize the vevs as:
\be
\langle \phi\rangle= v_\phi\,
\left(
\begin{array}{c}
\phi_1 \\
\phi_2
\end{array}
\right) \qquad 
\langle \xi\rangle= v_\xi\,
\left(
\begin{array}{c}
\xi_1 \\
\xi_2
\end{array}
\right) \qquad \langle \chi\rangle = v_\chi\qquad \langle \chi^\prime\rangle = v_\chi^\prime\,,
\ee
where $\phi_{1,2}$ and $\xi_{1,2}$ are adimensional quantities.
At the leading order we have:
\bea
w_d = g_1\,\xi_0\,\xi^2 + g_2\,\Delta_0\, \phi^2 + g_3\, \Delta_0\, \phi\,\chi\,.
\eea
The condition for the minima are:
\bea
\frac{\partial w_d}{\Delta_0^1} &=& g_2\,v_\phi^2\,\phi_1^2 + g_3\,v_\phi\,\phi_2\,v_\chi = 0 \nn
\\
&& \nn \\
\frac{\partial w_d}{\Delta_0^2} &=& g_2\,v_\phi^2\,\phi_2^2 + g_3\,v_\phi\,\phi_1\,v_\chi = 0
\\
&& \nn \\
\frac{\partial w_d}{\xi_0} &=& \xi_1\,\xi_2 = 0\,.\nn
\eea
The set of equations admit the solution:
\bea
\phi_1 = \phi_2 \qquad v_\chi=-\frac{g_2}{g_3}\,v_\phi\qquad \xi_1 = 0\,.
\eea
The relation among $v_\xi$ and $v_\phi$ allows us to assume a common
 order of magnitude for these vevs; on the other hand, 
the choice $\xi_1=0,\,\xi_2\ne0$ is not restrictive since the other one, with $\xi_1\ne0,\xi_2=0$ \cite{Feruglio:2007hi}, can be obtained 
acting with the generator $S$ on it.
At this stage, the flavon $\chi^\prime$ does not appear 
in the superpotential. However, the first corrections to $w_d$ involve $\chi^{\prime \,4}$ and read:
\bea
\delta w_d = g_1^\prime \,\xi_0\,\frac{\chi^{\prime \,4}}{\Lambda^2}\,.
\eea 
Since it appears with the driving field $\xi_0$, this term can modify the vev of the $\xi$ flavon; assuming a perturbed structure like:
\be
\langle \xi\rangle= v_\xi\,
\left(
\begin{array}{c}
\xi_1^\prime \\
1+\xi_2^\prime
\end{array}
\right) \,,
\ee
the minimizing equation (at the first order in the perturbations $\xi_{1,2}^\prime$) gives:
\bea
2\,g_1\,v_\xi^2\,\xi_1^\prime + g_1^\prime \,\left(\frac{v_{\chi^\prime}^2}{\Lambda}\right)^2=0\,.
\eea
Dividing this equation by $1/\Lambda^2$ and assuming, as usual, that $v_\xi/\Lambda \sim v_{\chi^\prime}/\Lambda\sim\varepsilon$, we 
can estimate $\xi_1^\prime\sim \varepsilon^2$ and also:
\bea
\frac{v_\chi^{\prime\,4}}{\Lambda^2} = - \frac{2\,g_1}{g_1^\prime}\,v_\xi^2\,\varepsilon^2.
\eea
The perturbation $\xi_2^\prime$ remains unspecified and we put it to zero.
Summarizing:
\be
\label{xicorr}
\langle \xi\rangle= v_\xi\,
\left(
\begin{array}{c}
\xi^\prime\,\varepsilon^2 \\
1
\end{array}
\right), \qquad v_\chi^{\prime}\sim v_\xi\,,
\ee
with $\xi^\prime$ a coefficient of ${\cal O}(1)$.
For the other two flavons $\phi$ and $\chi$, the first useful corrections arise at the level of five-flavon insertion. The corrective terms 
are then of relative ${\cal O}(\varepsilon^3)$ with respect to the leading order results and will be 
neglected in the following. Then, the NLO corrections to the mass matrices will be computed using the vev structures for $\phi$ and $\chi$ as 
given in eqs.(\ref{loflav})-(\ref{loflav2}).

\section{Next to leading order corrections}
\label{nlosect}
It is important to check that the previous results on the mixing angles are not destroyed once the 
corrections to the lagrangians are taken into account. 
\subsection{Charged leptons}
\label{chcorr}
The most relevant corrections come from the term:
\bea
\delta{\cal L}_{cl} = x_3\, L_e\, L^c\, h_d\, \frac{\phi\chi^\prime}{\Lambda^2}\,, 
\eea
which modifies the mass matrix as follows:
\be 
\label{chargedLO2}
\delta m_\ell= v_d\,
\left(\begin{array}{ccc}
0&x_3 \varepsilon^2& -x_3 \varepsilon^2\cr
0&0&0 \cr
0&0&0\end{array}\right)\,.
 \ee
Still working with real parameters for simplicity, we see that the expression of the masses are not modified 
whereas the mixing matrix is now given by:
\be 
U_\ell= 
\left(\begin{array}{ccc}
$\text{sign}$\left[(x_1-x_2)\,x_3\right]&\frac{\sqrt{2}\,x_3}{x_1-x_2}\,\varepsilon & 0\cr
\left|\frac{x_3}{x_1-x_2}\right|\,\varepsilon&\frac{1}{\sqrt{2}}&-\frac{1}{\sqrt{2}} \cr
-\left|\frac{x_3}{x_1-x_2}\right|\,\varepsilon&\frac{1}{\sqrt{2}}&\frac{1}{\sqrt{2}}\end{array}\right)\,.
 \ee
Then, there are new contributions to the leptonic $\theta_{12}$ and $\theta_{13}$ but not to the atmospheric
angle, since the $2-3$ sub-block  still remains unchanged at this order. 
The electron mass is not generated at this order but only at the level of 5-flavon insertion, from the 
following two (non-vanishing) operators:
\bea
L_e\,L_e^c\,h_d\,\left\{
\xi^2\,\phi^2\,\chi^\prime,\; \xi^2\,\phi\,\chi\,\chi^\prime\right\}\nn\,, 
\eea
which give a contribution of ${\cal O}(\varepsilon^5)$; reminding that $m_\tau \sim \varepsilon$,
the electron-to-tau mass ratio is of ${\cal O}(\varepsilon^4)$. Considering that this ratio should be some units of
$10^{-4}$, we deduce that $\varepsilon$ cannot be too different from $\approx 0.13$. 
It is important to stress, however, that this value of $\varepsilon$ is merely indicative,
as it can be enhanced or suppressed by a proper arrangement of the Yuwaka couplings. We will study more in details 
the magnitude of the breaking parameter in Sect.\ref{numan}.
As it will be clear in the next section, $U_\ell$ will provide the largest corrections to the LO results.

\subsection{Neutrinos}
At the next level of approximation, the light neutrino mass matrix can be still evaluated  using eq.(\ref{seesaw})
but expanded now at ${\cal O}(\varepsilon^2)$ with flavon alignments as discussed in Sect.\ref{flavonalignment}.
%
%
%
Here we want to comment that the filling of some of the vanishing 
entries in eq.(\ref{diracmass}) and (\ref{majo}) requires multiple-flavon insertions; for example, 
the elements (22), (23), (31) and (33) in $m_D$ are generated by operators with three flavons (like $L\,\nu^c\,\phi^3\,h_u$) and are
then of ${\cal O}(\varepsilon^3)$. 
All next-to-leading order Majorana operators are also of ${\cal O}(\varepsilon^3)$
since they contain at least four flavons. Then, the main contributions to the LO vanishing matrix elements in 
eq.(\ref{majo}) come from the vev shift in eq.(\ref{xicorr}) and are of ${\cal O}(\varepsilon^2)$.
From these considerations is not difficult to understand that the NLO light neutrino mass matrix 
is given by (still using $a=1$):
\be 
\label{massmu2}
m_\nu= -\frac{v_u^2}{v_\Phi}\,
\left(\begin{array}{ccc}
 0& \frac{b}{\beta}\,\varepsilon   & \frac{b \,\gamma}{\alpha\,\beta}\,\varepsilon^2 \cr
\frac{b}{\beta}\,\varepsilon  & -\frac{\delta}{\beta^2}\,\varepsilon & \left(\frac{\xi^\prime\beta^2-\gamma\delta}{\alpha\beta^2}\right)\varepsilon^2\cr
\frac{b\, \gamma}{\alpha\,\beta}\,\varepsilon^2& \left(\frac{\xi^\prime\beta^2-\gamma\delta}{\alpha\beta^2}\right)\varepsilon^2&\frac{1}{\alpha}
\end{array}\right)\,.
\ee
The expression of the neutrino mixing matrix is quite cumbersome; we have found that all eigenvalues and 
eigenvectors are corrected by (intricate) ${\cal O}(\varepsilon^2)$ terms. As previously stated, the charged lepton rotation $U_\ell$
gives the main corrections to the LO results for $\theta_{12}$ and $\theta_{13}$, whereas the atmospheric angle 
only receives ${\cal O}(\varepsilon^2)$ corrections from the neutrino sector.
The final results for the mixing angles are then:
\bea
\tan \theta_{12} &=& \tan \theta_{12}^{LO} + \frac{\sqrt{2}\,x_3\,(4b^2\beta^2+d^2)^{1/2}}{(x_1-x_2)
\sqrt{2b^2\beta^2+\delta[\delta+(4b^2\beta^2+d^2)^{1/2}]}}\,\varepsilon +{\cal O}(\varepsilon^2)\nn \\
\tan \theta_{23} &=& 1 +{\cal O}(\varepsilon^2) \label{finalres}\\
\sin \theta_{13} &=&\left|\frac{x_3}{x_1-x_2}\right|\varepsilon +{\cal O}(\varepsilon^2)\nn\,.
\eea
As we can see, it is not difficult to reconcile our results with the experimental data, barring accidental cancellations; 
in fact, the NLO
have preserved many good features of the LO results (large solar and atmospheric mixings) while 
producing a relatively large shift for $\theta_{13}\sim {\cal O}(\varepsilon)$.

\subsection{Effective mass terms and  $|m_{ee}|$}
The previous results are not substantially modified by effective Weinberg operators. 
Up to four-flavon insertion, the lagrangian is as follows:
\bea
{\cal L}_w=\frac{\alpha_w}{\Lambda_{LN}}\,L_e\,L_e\,h_u\,h_u + \frac{\beta_w}{\Lambda^3_{LN}}\,L\,L\,h_u\,h_u\,\xi^2 + 
\frac{\gamma_w}{\Lambda^4_{LN}}\,L_e\,L\,h_u\,h_u\,\xi^2\,\chi^\prime \,,
\eea
where $\Lambda_{LN}$ is the lepton number breaking scale.
The contributions to the neutrino mass matrix is then:
\be 
\label{massmu3}
m_w= \frac{v_u^2}{\Lambda_{LN}}\,
\left(\begin{array}{ccc}
\alpha_w&  -\gamma_w\,\frac{v_\Phi^3}{\Lambda_{LN}^3}  & 0 \cr
-\gamma_w\,\frac{v_\Phi^3}{\Lambda_{LN}^3}  & 0 & 0 \cr
0 & 0 &\beta_w\,\frac{v_\Phi^2}{\Lambda^2_{LN}}
\end{array}\right)\,,
\ee
so we still have a block-diagonal form.
In practise, the relevant contribution to $m_\nu$ is given by the first operator in ${\cal L}_w$, which fills the (11) vanishing entry in 
eq.(\ref{massmu2}) with a term of ${\cal O}(v_u^2/\Lambda_{LN})$. In the case $\Lambda_{LN}\sim \Lambda$, this term also 
contributes to  eq.(\ref{massesLO}) and eq.(\ref{solatmLO}), changing the coefficients in front of the $\varepsilon$ parameter but not
their order of magnitude. In the case $\Lambda_{LN}\gg \Lambda$, this term is negligible and the effective operators 
do not play any relevant role in the determination of the neutrino masses and mixings.
\noindent

To evaluate the prediction of our model for the effective mass $m_{ee}$, we work in the basis where the charged leptons are diagonal 
and extract the (11) matrix element of the rotated neutrino mass matrix, that is:
\bea
|m_{ee}| = |U_\ell^T\,m_\nu\,U_\ell|_{11}\,.
\eea
We get:
\bea
|m_{ee}| = \sin^2 \theta_{13}\, \left| \frac{ [2\, \alpha\,  b \,(x_1-x_2)+\beta \, x_3]}{\alpha\,  \beta \, x_3} \right|\,.
\eea
As expected in models for the normal hierarchy, $|m_{ee}|$ is small, at the level of $\theta_{13}^2$, although a clear correlation
with the reactor angle is lacking because of the ${\cal O}(1)$ coefficient.

\section{Numerical analysis of the lepton sector}
\label{numan}
\begin{figure}[h!]
\begin{center}
\includegraphics[scale=.425]{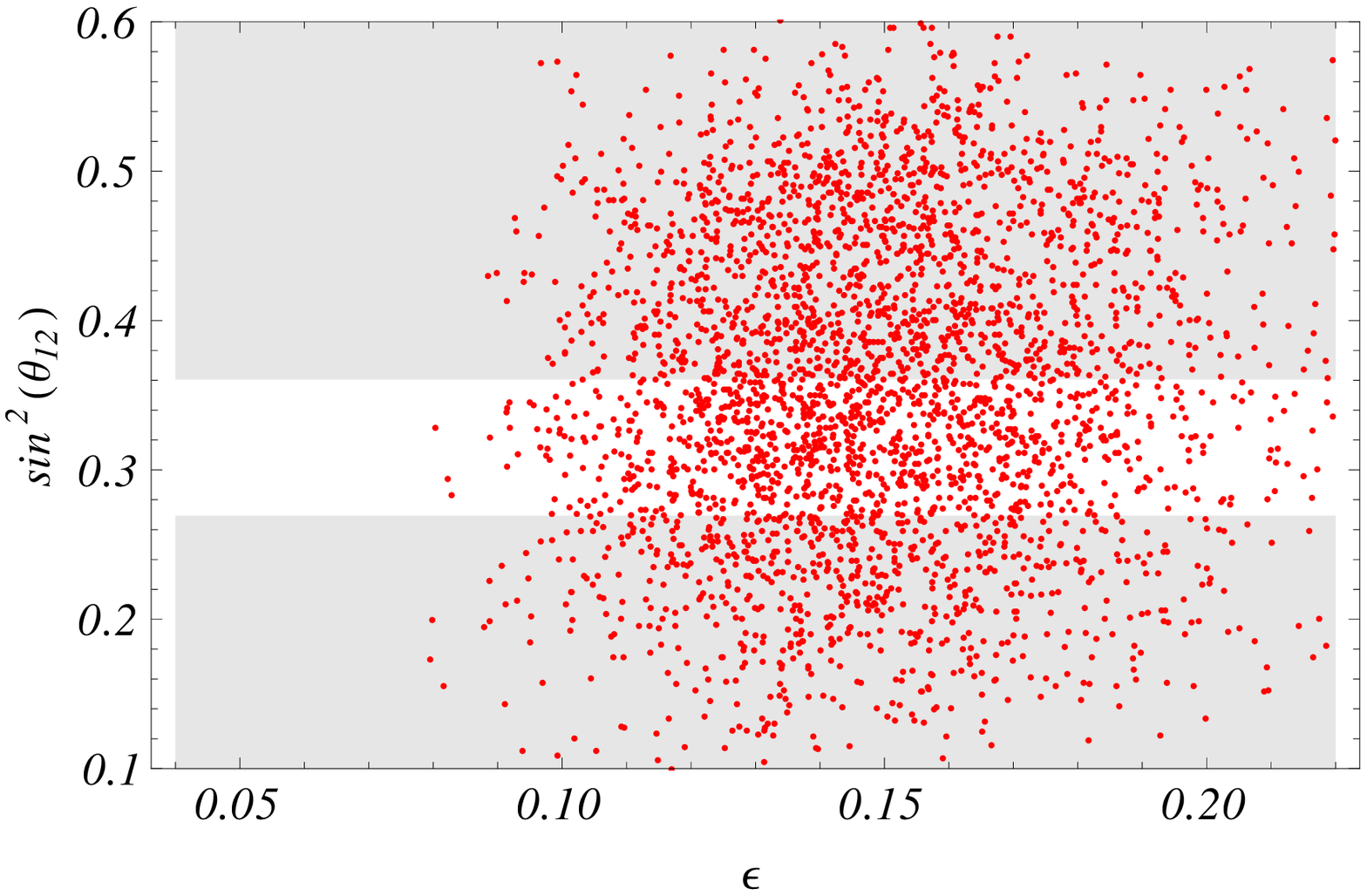}\qquad \includegraphics[scale=.425]{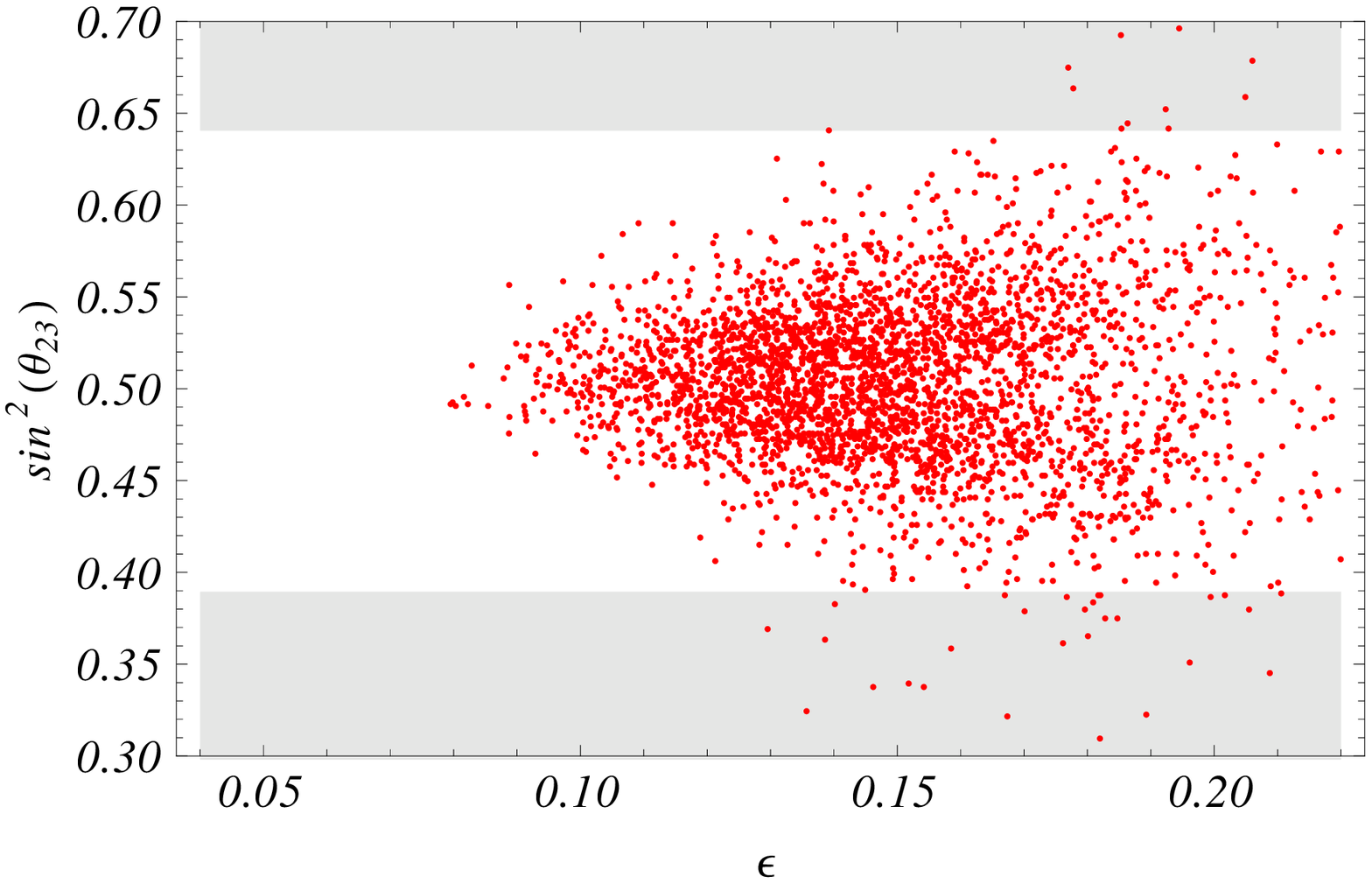}
\caption{\label{reference1} \it Mixing angles $\sin^2\theta_{12}$ (left panel) and $\sin^2\theta_{23}$ (right panel) 
as a function of the breaking scale $\varepsilon$.
The horizontal gray bands are the regions excluded by the experimental data at 3$\sigma$, as obtained from the second reference in \cite{Fogli:2011qn }.}
\end{center}
\end{figure}

\begin{figure}[h!]
\begin{center}
\includegraphics[scale=.425]{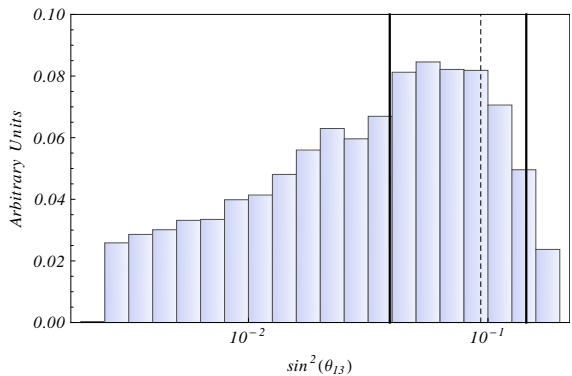}  
\caption{\label{reference2} \it Distribution of the valus of $\sin^2 2\theta_{13}$  as obtained from our model;
the vertical solid lines enclose the 3$\sigma$ range derived from eq.(\ref{res}) whereas the dashed line is the 
corresponding best fit point.}
\end{center}
\end{figure}

The main purpose of this section is to analyze in detail the implication of our model for the 
lepton masses and mixing. In doing that, we use the NLO charged lepton and neutrino mass matrices 
as given in eqs.(\ref{chargedLO})-(\ref{chargedLO2}) and eq.(\ref{massmu2}) and
perform a MonteCarlo simulation extracting complex lagrangian parameters with absolute values in the 
interval $[1/2,2]$ whereas the small $S_3$ breaking parameter is taken randomly in $[\lambda^2,\lambda]$. 
To study more in detail the magnitude of $\varepsilon$, we imposed {\it relaxed} bounds 
on the charged lepton mass ratios\footnote{More restricted bounds would only select a narrower range of $\varepsilon$'s.}:
\bea
\label{massratio}
2\times 10^{-3}<\frac{m_e}{m_\mu} < 9\times 10^{-3} \qquad 2\times 10^{-2}<\frac{m_\mu}{m_\tau} < 9\times 10^{-2}\,;
\eea
we also impose the constraints on the neutrino mass differences:
\bea
\label{massratio2}
7.09\times 10^{-5}<\Delta m^2_{sol} < 8.2\times 10^{-5} \,\text{eV}^2 \qquad 2.14\times 10^{-3}<\Delta m^2_{atm} < 2.76\times 10^{-3}\, \text{eV}^2\,,
\eea
derived from the second reference in \cite{Fogli:2011qn}, so that the value of $r$ is automatically reproduced.
The main results of such an analysis are presented in Figs.\ref{reference1}-\ref{reference2}. 
In the former, we are interested to the dependence of $\sin^2\theta_{12}$ and $\sin^2\theta_{23}$ on the breaking scale $\varepsilon$.
In these plots, the horizontal gray bands are the regions excluded by the experimental data at 3$\sigma$, as obtained from the second reference in \cite{Fogli:2011qn}.
We clearly see that $\varepsilon \sim {\cal O} (\lambda^2)$ are almost excluded because too small to fill the relations in eq.(\ref{massratio}). 
For every mixing angle, the bulk of the selected points is around $\varepsilon \sim 0.14$, not really different from the estimate we gave in 
Sect.\ref{chcorr}. The analytical results anticipated in eq.(\ref{finalres}) are also confirmed; in particular, we see that 
the solar angle (left plot) is mainly undetermined but a large fraction of the points fall 
into the allowed 3$\sigma$ range, as a consequence of the fact that the TBM approximation is, in some sense, contained in the model via the simple
relation in eq.(\ref{tbm}). For the atmospheric angle (right plot), the majority of the points are well inside the 3$\sigma$ range, showing also 
a tendency to a largest spread for $\varepsilon \gtrsim 0.14$, as indicated in eq.(\ref{finalres}).
Similar considerations could also be drawn for the dependence of the reactor angle on $\varepsilon$; however, insted of presenting a scatter plot,
we prefer to compare the $\sin^2 2\theta_{13}$ distribution as obtained from our numerical simulation directly with the Daya Bay result. This is shown in 
Fig.\ref{reference2}, where the 3$\sigma$ bounds derived from eq.(\ref{res}) are enclosed in the solid vertical lines, whereas the dashed line 
is the best fit point. We can appreciate that, although the distribution is quite broad,
the largest density of $\sin^2 2\theta_{13}$ extractions is just inside the allowed region. We stress that this result has been 
obtained with an {\it ab-initio} simulation of the charged lepton and neutrino mass matrices with no other constraint than 
those given in eqs.(\ref{massratio}-\ref{massratio2}). Notice that, given the large number of ${\cal O}(1)$ parameters, no definite predictions for the CP phase can be drawn. 

\section{The quark sector}
\label{quarksect}
\begin{figure}[h!]
\begin{center}
\includegraphics[scale=.9]{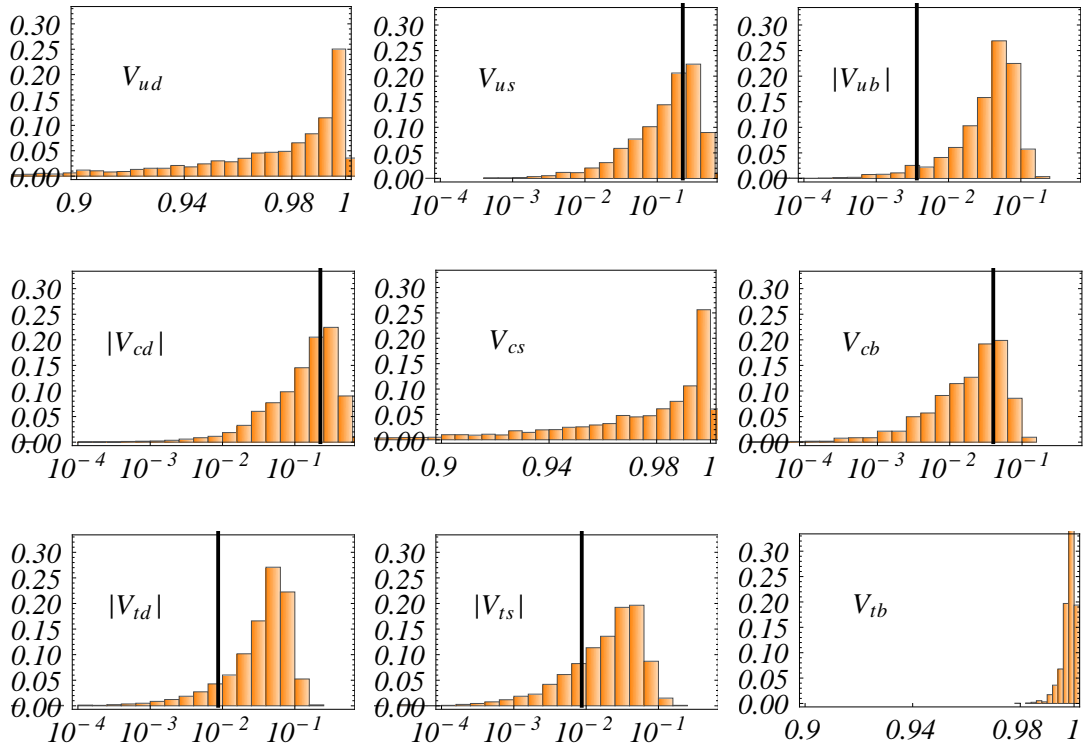} 
\caption{\label{ckmstructure} \it Distributions of the CKM matrix elements as obtained in our model. The best fit values \cite{utfit} of the off-diagonal matrix 
elements are shown with a solid vertical line.}
\end{center}
\end{figure}

The $S_3 \times Z_3 \times Z_6$ symmetry  provides a good description of the quark sector also; we use the same flavon fields and alignments described 
in the previous sections. The first two families of left-handed quarks are assigned to a $2$ representation of $S_3$ whereas all other 
fields belong to singles $1$ or $1_2$ in the case of $b^c$ (see Tab.\ref{transformq}).
\begin{table} [h!]
\centering
\begin{tabular}{|c||c|c|c|c|c|c|c|c|}
\hline
{\tt Field}& $Q_L$ & $q_{3_L}$ & $u^c$ & $c^c$ & $t^c$ & $b^c$ & $s^c$& 
$d^c$   \\
\hline
$S_3$ & $2$ & $1$ & $1$ & $1$ & $1$ & $1_2$ &$1$ &$1$   \\
\hline
$Z_6$ & $\omega^5$ & $1$ & $\omega^4$ & $1$ & $1$ & $\omega$ & 1 &
$\omega^4$   \\
\hline
$Z_3$ & $\omega$ & 1 & $1$ & $\omega^2$ & 1 & $1$ & $\omega^2$ &
1  \\
\hline
\end{tabular}
\caption{\it Transformation properties of quarks under $S_3 \times Z_3 \times Z_6$.}
\label{transformq}
\end{table}
The lagrangian in the up-quark sector, including all relevant operators to generate non-vanishing entries in the mass matrix, 
reads as follows:
\bea
{\cal L}_{up} &=&a_u \, q_{3_L}\,h_u\,t^c + b_u\, q_{3_L}\,h_u\,c^c\,\frac{\xi^2\,\phi}{\Lambda^3} + 
c_u\, q_{3_L}\,h_u\,u^c\,\frac{\xi^2}{\Lambda^2}+d_u\, Q_{L}\,h_u\,t^c\,\frac{\phi^2\,\chi^\prime}{\Lambda^3}\nn 
\\ & & \nn \\
&&+e_u\, Q_{L}\,h_u\,c^c\,\frac{\xi^2\,\chi^\prime}{\Lambda^3}+e_u^\prime\, Q_{L}\,h_u\,c^c\,\frac{\xi\,\chi^{\prime 3}}{\Lambda^4}+
f_u\, Q_{L}\,h_u\,u^c\,\frac{(\xi^2)_2\,(\phi^2)_2\,\chi^{\prime}}{\Lambda^5} \label{lagup} 
\\ & & \nn \\
&&+f_u^\prime\, Q_{L}\,h_u\,u^c\,\frac{(\xi^2)_2\,(\phi^2)_1\,\chi^{\prime}}{\Lambda^5}+
f_u^{''}\, Q_{L}\,h_u\,u^c\,\frac{(\xi^2)_2\,\phi\,\chi\,\chi^{\prime}}{\Lambda^5}+
f_u^{'''}\, Q_{L}\,h_u\,u^c\,\frac{(\xi^2)_2\,\chi^2\,\chi^{\prime}}{\Lambda^5}\nn\,,
\eea
where we have indicated the $S_3$ contractions when necessary. The related mass matrix is as follows:
\bea 
\label{massup}
m_{up}= v_u\,
\left(\begin{array}{ccc}
(f_u +f_u^{''})\, \varepsilon^5& -e_u^\prime\,\varepsilon^4 & -d_u\,\varepsilon^3 \cr
-(2 f_u^\prime +f_u^{'''})\, \varepsilon^5 & e_u\,\varepsilon^3  & d_u\,\varepsilon^3 \cr
2\,c_u\,\xi^\prime\, \varepsilon^4 & b_u\,\varepsilon^3 & a_u
\end{array}\right)\,.
\eea
First of all, we observe that the mass hierarchy is well reproduced for the same value of $\varepsilon$ as deduced
from the lepton sector; in particular, we have:
\bea
m_{u} = v_u\,(f_u +f_u^{''})\, \varepsilon^5 \qquad m_c = v_u\,e_u\,\varepsilon^3 \qquad m_t=v_u\,a_u\,.
\eea
Then, taking real Yukawas for simplicity, the matrix $U_{up}$ diagonalizing $m_{up}\,m_{up}^\dagger$ is given by:
\bea 
\label{Uup}
U_{up}= 
\left(\begin{array}{ccc}
1& \frac{e_u^\prime}{e_u}\,\varepsilon & \frac{d_u}{a_u}\,\varepsilon^3 \cr
-\frac{e_u^\prime}{e_u}\,\varepsilon & 1  & -\frac{d_u}{a_u}\,\varepsilon^3 \cr
-\frac{d_u}{a_u}\,\varepsilon^3  & \frac{d_u}{a_u}\,\varepsilon^3&1 
\end{array}\right)\,.
\eea
This matrix goes into the desired direction: at LO, $U_{up}$ is the identity and at the NLO
there exist a well defined hierarchy among the (12) element, not far away from the value of the Cabibbo angle, 
and the other off-diagonal matrix elements, which contribute to $V_{cb}$ and $V_{ub}$. 
From the charge assignment in Tab.\ref{transformq}, we see that the operators 
in the down sector involving $d^c$ and $s^c$ are the same as those with  $u^c$ and $c^c$, respectively (with the obvious replacement 
$h_u \to h_d$), whereas all operators including $b^c$  do not have a corresponding in ${\cal L}_{up}$; 
then we have:
\bea
{\cal L}_{down} &=&a_d \, q_{3_L}\,h_d\,b^c\,\frac{\chi^\prime}{\Lambda}  + b_d\, q_{3_L}\,h_d\,s^c\,\frac{\xi^2\,\phi}{\Lambda^3} + 
c_d\, q_{3_L}\,h_d\,d^c\,\frac{\xi^2}{\Lambda^2}+d_d\, Q_{L}\,h_d\,b^c\,\frac{(\phi^2)_1\,\xi}{\Lambda^3}\nn 
\\ & & \nn \\
&&+d_d^\prime\, Q_{L}\,h_d\,b^c\,\frac{(\phi^2)_2\,\xi}{\Lambda^3}+d_d^{''}\, Q_{L}\,h_d\,b^c\,\frac{\chi^2\,\xi}{\Lambda^3}
+d_d^{'''}\, Q_{L}\,h_d\,b^c\,\frac{\chi\,\phi\,\xi}{\Lambda^3} \\ && \nn \\&&
+e_d\, Q_{L}\,h_d\,s^c\,\frac{\xi^2\,\chi^\prime}{\Lambda^3}+e_d^\prime\, Q_{L}\,h_d\,s^c\,\frac{\xi\,\chi^{\prime 3}}{\Lambda^4}+
f_d\, Q_{L}\,h_d\,d^c\,\frac{(\xi^2)_2\,(\phi^2)_2\,\chi^{\prime}}{\Lambda^5} \nn 
\\ & & \nn \\
&&+f_d^\prime\, Q_{L}\,h_d\,d^c\,\frac{(\xi^2)_2\,(\phi^2)_1\,\chi^{\prime}}{\Lambda^5}+
f_d^{''}\, Q_{L}\,h_d\,d^c\,\frac{(\xi^2)_2\,\phi\,\chi\,\chi^\prime}{\Lambda^5}+
f_d^{'''}\, Q_{L}\,h_d\,d^c\,\frac{(\xi^2)_2\,\chi^2\,\chi^{\prime}}{\Lambda^5}\nn\,,
\eea
with mass matrix as:
\bea 
\label{massdrr}
m_{down}= v_d\,\varepsilon\,
\left(\begin{array}{ccc}
(f_d +f_d^{''})\, \varepsilon^4& -e_d^\prime\,\varepsilon^3 & -(2 d_d + d_d^{''})\,\varepsilon^2 \cr
-(2 f_d^\prime + f_d^{'''})\,\varepsilon^4 & e_d\,\varepsilon^2  & (d_d^\prime + d_d^{'''})\,\varepsilon^2 \cr
2\,c_d\,\xi^\prime\, \varepsilon^3 & b_d\,\varepsilon^2 & a_d
\end{array}\right)\,.
\eea
Again, the mass ratios are well reproduced, since:
\bea
m_{d} = v_d\,(f_d +f_d^{''})\, \varepsilon^5 \qquad m_s = v_d\,e_d\,\varepsilon^3 \qquad m_b=v_d\,a_d\,\varepsilon\,.
\eea
Also, the bottom-to-top mass ratio is given by 
\bea
\frac{m_b}{m_t} = \left(\frac{a_d}{a_u}\right)\,\varepsilon\,(\tan \beta)^{-1}\,,
\eea
implying $\tan \beta \sim 5$ for Yukawas of ${\cal O}(1)$.
The matrix $U_{down}$ diagonalizing $m_{down}\,m_{down}^\dagger$ is given by:
\bea 
\label{Ud}
U_{d}= 
\left(\begin{array}{ccc}
1& \frac{e_d^\prime}{e_d}\,\varepsilon & (2 d_d + d_d^{''})\,\varepsilon^2 \cr
-\frac{e_d^\prime}{e_d}\,\varepsilon & 1  & -(d_d^\prime + d_d^{'''})\,\varepsilon^2 \cr
-(2 d_d + d_d^{''})\,\varepsilon^2   & (  d_d^\prime + d_d^{'''})\,\varepsilon^2&1 
\end{array}\right)\,.
\eea
Two comments are in order: on the one-hand, the $(12)$ element is still of the correct order of magnitude 
to explain, in combination with the result from the up sector, the value of the Cabibbo angle.
On the other hand, all other off diagonal entries are smaller that $(12)$ but the (13) element is a bit larger than
the required values to fit $V_{ub}$. In fact, these elements in eq.(\ref{Ud}) are the dominant 
contributions to $V_{cb}$ and $V_{ub}$ in the CKM, since the corresponding matrix elements in 
eq.(\ref{Uup}) are generally smaller. In fact, we get:
\bea
V_{us} &=& \left(\frac{e_d^\prime}{e_d}-\frac{e_u^\prime}{e_u}\right)\,\varepsilon \qquad V_{ub}=(2 d_d + d_d^{''})\,\varepsilon^2 \qquad
V_{cb} =-(d_d^\prime + d_d^{'''})\,\varepsilon^2\nn \\  &&\\ \nn
V_{cd} &=& - V_{us} \qquad V_{td} = - V_{ub} \qquad V_{ts} = - V_{cb}\,.
\eea
Obviously, the last equalities are lifted once the ${\cal O}(\varepsilon^3)$ terms are taken into account.
To corroborate the previous considerations, we perform a numerical simulation of the up and down
mass matrices of eqs.(\ref{massup}) and (\ref{massdrr}) extracting, as we did for the lepton sector,
complex Yukawa parameters with absolute values in the 
interval $[1/2,2]$. We fixed the breaking scale $\varepsilon = 0.15$, as suggested by the same procedure
done in the lepton sector.
We also impose the following constraints on the relevant mass ratios \cite{pdg}:
\bea
5.40\times 10^{-6}<\frac{m_{u}}{m_{t}}<1.18\times 10^{-5}   &\qquad& 3.60\times 10^{-3}<\frac{m_{c}}{m_{t}}<4.10\times 10^{-3} \nn
\\ && \\
 7.90\times 10^{-4}<\frac{m_{d}}{m_{b}}<1.34\times 10^{-3}   &\qquad& 1.60\times 10^{-2}<\frac{m_{s}}{m_{b}}<2.90\times 10^{-2} \nn\,.
\eea
Our results for every CKM matrix elements are shown in fig.\ref{ckmstructure}, where we plot the absolute values of the nine 
distributions of the $V_{ij}$ entries. The diagonal entries are displayed in linear scale whereas we adopt a log scale for the 
off-diagonal elements; for them, we also showed the best fit values \cite{utfit} with a solid vertical line.
As we can see, it is very easy in our model to reproduce, with {\it no fine-tuning}, 
the correct values of $V_{us},V_{cd}$ as well as $V_{cb},V_{ts}$ while a small discrepancy remains 
with the best fit of $V_{ub}$ and $V_{td}$, as anticipated above. To make also these matrix elements fully compatible with the data,
we need a moderate cancellation among the Yukawas $d_d^\prime $ and $d_d^{'''}$, see eq.(\ref{Ud}).
We do not present any plot related to the CP phase since the large number of ${\cal O}(1)$ parameters does 
not allow any definite prediction.
As a last comment, we observe that taking a charge assignment for the field $b^c$ as the one adopted for $t^c$, we would get 
a structure of $U_d$ similar to $U_{up}$ as given is eq.(\ref{Uup}); this automatically would imply 
$V_{ub}\sim V_{cb}\sim V_{td}\sim V_{ts}\sim \varepsilon^3$, so an enhancement should be invoked to reproduce the values 
of  $V_{ub}$ and $V_{td}$. In this case, the bottom-to-tau mass ratio would be completely explained by large
$\tan \beta$.

\section{Conclusions}
The Daya Bay Collaboration has released the measurement of the reactor angle $\theta_{13}$, showing a 5.2$\sigma$ 
discrepancy from zero. From the model building point of view, neutrino mass textures 
predicting a vanishing $\theta_{13}$ at leading order seem to be less appealing, unless 
large corrections bring the reactor angle to values compatible with the recent results, without destroying the 
predictions for the other mixing parameters. 
In this paper we have presented a see-saw SUSY model for fermion masses and mixing based on the non-abelian discrete symmetry $S_3$, whose 
main result is the prediction of a large $\theta_{13}\sim 0.13$, fully 
compatible with the Daya Bay claim of eq.(\ref{res}). 
Other remarkable  features of our construction are:
\begin{itemize}
\item in the lepton sector, the spectrum is of normal type, with $\theta_{12}$ and $\theta_{23}$
compatible with their experimental allowed ranges;
\item in the quark sector, we obtained a good description of the relevant mass ratios and the absolute values of
all the CKM matrix elements (including the Cabibbo angle) but $V_{ub},V_{td}$, for which we
need a moderate fine-tuning among the Yukawas defining these matrix elements;
\item the flavon alignments needed to get the previous results 
are natural minima of the superpotential in the SUSY limit.
\end{itemize} 

\section*{Acknowledgments} 
We thank Guido Altarelli for some interesting comments and discussions. 
We also acknowledge  MIUR (Italy) for financial support
under the program "Futuro in Ricerca 2010 (RBFR10O36O)`` and CERN, where
this work was conceived.

\end{document}